\documentstyle[12pt,epsf]{article}
\def\beq{\begin{equation}}
\def\eeq{\end{equation}}
\def\beqn{\begin{eqnarray}}
\def\eeqn{\end{eqnarray}}
\def\lsim{\mathrel{\rlap{\lower3pt\hbox{\hskip0pt$\sim$}}
    \raise1pt\hbox{$<$}}}
\def\gsim{\mathrel{\rlap{\lower4pt\hbox{\hskip1pt$\sim$}}
    \raise1pt\hbox{$>$}}}

\begin{document}
\renewcommand{\theequation}{\thesection.\arabic{equation}}
\begin{titlepage}

\begin{flushright}
TPI-MINN-02/47, UMN-TH-2120/2\\
ITEP-TH-21/03, CERN-TH/2003-071
\end{flushright}

\vspace{0.3cm}

\begin{center}
\baselineskip25pt
{\Large\bf {\boldmath${\cal N}=2$} Sigma Model with Twisted Mass and
Superpotential: Central Charges and Solitons}

\vspace{0.3cm}
{ \bf  A. Losev$^{a,b}$} and { \bf M.~Shifman$^{b,c}$}

{\em $^{a}\,$Institute of Theoretical and Experimental Physics,}\\
\vspace{-0.35cm}
{\em B. Cheremushkinskaya 25, Moscow
117259, Russia$^\star$}

{\em $^{b}\,$William I. Fine Theoretical Physics Institute, University of Minnesota,}\\
\vspace{-0.35cm}
{\em 116 Church St. S.E., Minneapolis, MN 55455, USA$^\star$}

{\em ${}^c\,$Theory Division, CERN}\\
\vspace{-0.35cm}
{\em CH-1211 Geneva 23, Switzerland}

\end{center}

\vspace{1cm}

\begin{center}

{\large\bf Abstract} \vspace*{.25cm}

\end{center}

We consider supersymmetric sigma models on the K\"ahler target 
spaces, with 
twisted mass. The K\"ahler spaces are assumed to have holomorphic
Killing vectors.
 Introduction of a superpotential of a special type is known
to be consistent with ${\cal N}=2$ superalgebra (Alvarez-Gaum\'{e} and Freedman).
We show that the algebra acquires 
{\em central charges} in the anticommutators
$\{ Q_L, Q_L\}$ and $\{ Q_R, Q_R\}$.
These central charges have no parallels,
and they can exist only in two dimensions.
The central extension of the ${\cal N}=2$ superalgebra
we found paves the way to a novel phenomenon ---
spontaneous breaking of a part of supersymmetry.
In the general case 1/2 of supersymmetry is spontaneously
broken (the vacuum energy density is positive),
 while the remaining 1/2 is realized linearly. 
In the model at hand the standard fermion number is not defined,
so that   the Witten index as well as the
Cecotti-Fendley-Intriligator-Vafa index are useless.
We show how to construct an index  for 
counting short multiplets in internal algebraic terms
which is well-defined in spite of the absence of the standard fermion number.
Finally, we outline derivation of the {\em quantum anomaly} in
the anticommutator $\{ \bar Q_L, Q_R\}$.

\vspace{0.5cm}

\noindent
\rule{2.4in}{0.25mm}\\
$^\star$ Permanent address.

\end{titlepage}

\newpage

\section{Introduction}
\label{intro}
\setcounter{equation}{0}

Supersymmetric ${\cal N}=2$ two-dimensional
sigma models with twisted mass term
possess a variety of central charges and exhibit (see e.g. \cite{Dorey})
nontrivial phenomena similar to those of the Seiberg-Witten theory in four
dimensions.  It was noted long ago \cite{Alvarez}
that   further deformation of
this model by a judiciously chosen superpotential is possible.
In this paper we show that such deformation
results in emergence of a  central charge of a novel type.
(A similar algebra was considered previously \cite{LSQM}
in a quantum-mechanical context.)  As a consequence,
${\cal N}=2$ supersymmetry is {\em spontaneoulsy} broken down to
${\cal N}=1$, in defiance of a well-known theorem forbidding this phenomenon.
Moreover,
Witten's index \cite{WitIndex}
turns out to be useless  for counting the
number of the {\em vacuum} states in this case, since it
always vanishes. The Cecotti-Fendley-Intriligator-Vafa (CFIV)
index \cite{CFIV},
$
I_{\rm CFIV} = {\rm Tr}\, F(-1)^F\,,
$
which might be appropriate for counting vacua
(rather than BPS solitons!),
does not work in this model either, since
the standard definition of the fermion number does not work.
We show how   new indices, ${\cal I}_{2,1/2}$ and ${\cal I}_{2,1/4}$
(formulated in internal
algebraic terms) replacing that of CFIV, can be introduced
following the line of reasoning of Ref. \cite{LSV}.
The index ${\cal I}_{2,1/2}$ counts the number of 1/2 BPS-saturated vacua,
while ${\cal I}_{2,1/4}$ counts the number of 1/4 BPS-saturated
kinks.

In the simplest example including twisted mass and superpotential
we find explicit 1/4 BPS kink solutions. Derivation of anomaly
in the  central charges is outlined. 

\section{Sigma model and  twisted mass}
\label{twistedmass}
\setcounter{equation}{0}

 Two-dimensional sigma models with twisted mass
were first constructed in Ref.~\cite{Alvarez}. The superspace/superfield
description was developed in Refs.~\cite{Gates:1983py,Gates:nk}.
In particular, the notion of a twisted  chiral superfield
was introduced in the second of these works\,\footnote{The word ``twisted'' 
appears for the first time in the given context in Ref.~\cite{Gates:nk}.}.
Our task in this section is to briefly review
introduction  of the twisted mass $m$
in the sigma model of a general form and
set the notation to be used throughout the paper.

For   K\"{a}hler target space endowed with the metric
$G_{\bar j\, i}$ the Lagrangian of the undeformed (i.e. $m=0$) model is
\beqn
{\cal L}_0 =
G_{\bar j\, i}\, \partial^\mu\bar\phi^{\,\bar j}\, \partial_\mu\phi^i
+\frac{i}{2} G_{\bar j\, i}\,
\bar\Psi^{\bar j} \!\stackrel{\leftrightarrow}{\not\! \! D}\Psi^{i}
+R_{\bar j i k\bar{l}}\,
\bar\Psi_L^{\,\bar j}\Psi_L^i\Psi_R^k\bar\Psi_R^{\,\bar l}
\,,
\label{one}
\eeqn
where  $D_\mu$ is the covariant derivative,
\beq
 D_\mu\,\Psi^i = \partial_\mu\,\Psi^i +
\partial_\mu\phi^k\, \Gamma_{kl}^i\,\Psi^l
\,.
\label{two}
\eeq
Here $\Gamma$'s stand for the
Christoffel symbols,  $R_{\bar{j} i k\bar{l}}$
is the curvature tensor, while $\Psi$ denotes a two-component spinor,
\beq
\Psi =\left(\begin{array}{cc}
\Psi_R \\
\Psi_L
\end{array}
\right),
\eeq
so that the kinetic term of the fermions can be identically rewritten as
\beq
\frac{i}{2} G_{i\bar j}\,
\bar\Psi^{\bar j} \!\stackrel{\leftrightarrow}{\not\! \! D}\Psi^{i}
=
\frac{i}{2}
G_{i\bar j}\left\{\bar\Psi_L^{\bar j}
  \!\stackrel{\leftrightarrow}{\not\! \! D}_R\Psi_L^i
+ \bar\Psi_R^{\bar j}  \!\stackrel{\leftrightarrow}{\not\! \! D}_L\Psi_R^i
\right\}\,.
\label{three}
\eeq
The right and left derivatives are defined in Eq.~(\ref{rlder}).

In addition, one can introduce a
$\theta$ term which has the following general form
\beq
{\cal L}_\theta =\frac{i\theta}{2\pi}\, B_{i\bar j}\, \varepsilon^{\mu\nu}
\partial_\mu \bar\phi^{\, \bar j}\partial_\nu\phi^i \,,
\label{four}
\eeq
where $B_{i\bar j}$ is a
closed 2-form (i.e. $\partial_k B_{i\bar j}=\partial_i B_{k\bar j}$
and h.c.). For the particular example of the CP(1) model, to be discussed
in some detail below, $B_{i\bar j}$
can be chosen proportional to the round metric
$G_{i\bar j} =\delta_{i\bar j} (1+|\phi|^2)^{-2}$.

If the target space of the ${\cal N}=2$ sigma model (\ref{one}) has isometries,
one can carry out
\cite{Alvarez}  a mass deformation of the model, by introducing the
so-called twisted mass,
without breaking ${\cal N}=2$.
Suppose $V=\{v^i\,,\,\, \bar v^{\,\bar j}\}$ is a
holomorphic vector field, i.e. $v^i = v^i (\phi)$,
$\bar v^{\,\bar j} =\bar v^{\,\bar j} (\bar \phi) $ and the Lie derivative ${\cal L}_V$
annihilates the metric,
\beqn
{\cal L}_V G_{\bar j\, i}
&\equiv& {\cal L}_v G_{\bar j\, i} +
{\cal L}_{\bar v} G_{\bar j\, i }=0\,;\nonumber\\[2mm]
{\cal L}_v G_{\bar j\, i}
&\equiv& v^k\partial_k G_{\bar j\, i} +
\partial_i\, v^k G_{\bar j\, k}\,;\nonumber\\[2mm]
{\cal L}_{\bar v} G_{\bar j\, i}
&\equiv& \bar v^{\,\bar k}\partial_{\bar k}G_{\bar j\, i}+
\partial_{\bar j}\bar v^{\,\bar k}G_{\bar k\, i}
 \,.
\label{five}
\eeqn
Then the twisted mass term takes the form
\beq
{\cal L}_m=  \left(  m\bar m\right)  \,
G_{ \bar j\, i } \,  \bar v^{\,\bar j}\, v^i
- m\left(  {\cal L}_vG_{\bar j\, i}\right)
\bar\Psi_L^{\,\bar j}\Psi_R^i +\bar m
 \left( {\cal L}_{\bar v} G_{\bar j\, i} \right)  \bar\Psi_R^{\,\bar j}\Psi_L^i
\,,
\label{six}
\eeq
where $m$ is a complex parameter.
Note that Eq. (\ref{six}) is invariant under vector rotations of the fermion
fields, $\Psi_{L,R}\to \exp (i\alpha)\Psi_{L,R}$.
The corresponding charge can be viewed as a
fermion number. The axial transformation of the fermion fields,
$\Psi_{L}\to \exp (i\alpha)\Psi_{L}$,
$\Psi_{R}\to \exp (-i\alpha)\Psi_{R}$, allows one
to  rotate away  the phase of $m$ at a price of redefining the $\theta$ angle.

The easiest way to derive Eq. (\ref{six}) is to choose the coordinate frame
$\Phi^j$ ($j=0,1,...,n$) in
such a way that 
\beq
v^j = iv_0\, \delta^{0j} \,,\qquad v_0={\rm real\,\,\,  const}\,.
\label{seven}
\eeq
Below the coordinates $\Phi_0,\,\,\bar\Phi_0$ will be referred to
as special.
With this choice, the K\"{a}hler term
\beq
\int d^4\theta K\left(\Phi_0 +\bar\Phi_0 +2\bar m\theta_L\bar\theta_R +
2 m \theta_R\bar\theta_L
\right)
\label{eight}
\eeq
leads to ${\cal N}=2$ preserving sigma model with the twisted mass.

The choice of the coordinates in Eqs. (\ref{one})
and (\ref{six}) is different from the special one.
Here   the coordinates $\phi\,,\,\,\bar\phi$
are such   that the K\"{a}hler potential $K$ depends on the
product $\bar\phi\phi$, so that the model is invariant
under phase rotations\footnote{In what follows we will limit ourselves to the
K\"{a}hler potentials depending on
$\bar\phi \phi$ so that U(1) is guaranteed to be isometry of the metric.}
\beq
\phi\to \phi \, e^{i\alpha }\,,\qquad \bar\phi \to \bar\phi\, e^{-i\alpha }\,.
\label{nine}
\eeq
The special coordinates are related to
$\phi\,,\,\,\bar\phi$ through the transformation 
$$
\phi = e^{\Phi_0}\,, \quad \phi_j = \Phi_j\,,\quad j >0
\,.$$

\section{Introduction of  superpotential}
\label{introsup}
\setcounter{equation}{0}

The question is whether one can introduce a superpotential in the
model with the twisted mass without breaking ${\cal N}=2$.
It is clear that a generic superpotential destroys isometry of the target space;
in this case combining the twisted mass with a superpotential
breaks ${\cal N}=2$ {\em explicitly}.
One can find, however, a special superpotential \cite{Alvarez}
preserving the isometry. In the special coordinates it has the form
\beq
d {\cal W} = w\, d\Phi_0\,,
\label{ten}
\eeq
($w$ is a complex constant)
which is invariant under shifts of $\Phi_0$ (cf. Eq. (\ref{seven})).
Then in the coordinates used in   Eqs. (\ref{one})  and (\ref{six})
\beq
d {\cal W} = w\, \frac{d\phi}{\phi}\,.
\label{eleven}
\eeq
Note that the superpotential ${\cal W}= w\,\ln\phi $ is multivalued --
 this feature is unavoidable. This causes no problems, however, since
$d{\cal W}$ is well-defined, and this is all we need.
It is worth mentioning 
that such multivalued superpotentials appeared previously \cite{Hori:2000kt}
in the context of the K\"{a}hler sigma models in the
mirror descriptions of the 
projective spaces with twisted mass term.

The superpotential enters through   $F$ term. This generates the following
additional term in the Lagrangian
\beq
{\cal L}_F = - G^{\, i\, \bar j} \left(
\partial_i {\cal W}\right) \left( \partial_{\bar j}\,\bar {\cal W}\right)
-\left\{ \left(D_i \partial_j{\cal W}\right)\Psi_R^j
\Psi_L^i +\mbox{h.c.}\right\}\,.
\label{twelve}
\eeq
Assembling Eqs.   (\ref{one}), (\ref{six})  and (\ref{twelve})
we get the full Lagrangian of the deformed model. 

Note that the second term
in Eq. (\ref{twelve}) forbids the vector
transformation of the fermion field.
On the other hand, the fermion mass term in Eq. (\ref{six})
forbids the axial transformation.
Thus, the
standard fermion number is {\em not} defined in the model including both
twisted mass term and   superpotential.
A U(1) symmetry involving both fermion and boson fields does exist, however;
the Lagrangian  (\ref{one}), (\ref{six})  and (\ref{twelve})
is invariant under
\beqn
\Psi_{L,R} &\to& \Psi_{L,R}\,
e^{i\alpha}\,,\qquad \phi \to \phi \, e^{i\alpha}\,;
\nonumber\\[2mm]
\bar\Psi_{L,R} &\to& \bar\Psi_{L,R}\, e^{-i\alpha}\,,
\qquad \bar\phi \to \bar\phi \, e^{-i\alpha}\,.
\label{thirteen}
\eeqn
The corresponding conserved current is\,\footnote{Equation (\ref{fourteen}) 
in the given form is in fact valid only
in the case of {\em one} complex coordinate $\phi$ 
and, correspondingly, {\em one} $\Psi$, so that all 
target space indices $i,\,\,\bar{j}$, an so on,
might  be omitted.
 }
\beqn
{\cal J}^\mu_{\rm U(1)} & =&  G_{\bar j \, i }\left( \bar\phi^{\bar j}
\, i \stackrel{\leftrightarrow}{\partial}^{\, \mu}\!\phi^i +
\bar\Psi^{\bar j}\gamma^\mu\Psi^i
\right)+ \frac{1}{2}
\left( \bar{\phi}^{\bar{i}}\bar{\Psi}^{\bar{j}}\gamma^{\mu}
\Psi^i \bar\Gamma_{\bar{i}\bar{j}}^{\bar{k}}G_{i\bar{k}}+
 \phi^{i} \bar\Psi^{\bar i}\gamma^{\mu}
{\Psi}^{j} \Gamma_{ij}^{k}G_{k\bar{i}}\right)
 ,
\nonumber\\[3mm]
\qquad q_{\rm U(1)}&\equiv&  \int dx {\cal J}^0_{\rm U(1)}\,.
\label{fourteen}
\eeqn

\section{Superalgebra}
\label{salgebra}
\setcounter{equation}{0}

As was mentioned,
the model given by Eqs. (\ref{one}), (\ref{six}), (\ref{twelve})
is ${\cal N}=2$ supersymmetric, i.e. possesses four conserved supercharges.
The superalgebra emerging after inclusion of the superpotential is
rather peculiar, however: it contains a central charge
of the type which seems to escape theorists' attention previously.
Here we will  discuss it in some detail.

The easiest way to derive the corresponding  superalgebra
at  tree level (we will   discuss possible anomalies in Sect. \ref{anomr})
is to consider the model in the flat limit, i.e. $G_{\bar j\, i} =1$.
Moreover, for our 
present purposes we may drop target space indices ${\bar j\,,\, i}$
limiting ourselves to one (complex) field.

Then the conserved supercharges are\footnote{It is, perhaps, more convenient
to give supercurrents in the following concise form
$J_R^+ =(\partial_R\bar\phi)\Psi_R$, $J_R^- =
i \bar m\bar\phi \Psi_L -i (\bar w/\bar \phi )\bar\Psi_L$,
$J_L^+ = im\bar\phi\Psi_R + i (\bar w/\bar\phi )\bar\Psi_R)$,  $J_L^-=
(\partial_L\bar\phi )\Psi_L$ where $J^{\pm} =(1/2)(J^0\pm J^1)$.}
\beqn
Q_R &=&\int dx \left\{\left(
\partial_R\bar \phi \right)\Psi_R + i \bar m\bar\phi\Psi_L
-i\frac{\bar w}{\bar\phi}\bar\Psi_L
\right\}\,,\nonumber\\[2mm]
\bar Q_R &=&\int dx \left\{\left(
\partial_R\phi \right)\bar \Psi_R - i m\phi\bar \Psi_L
+i\frac{ w}{\phi}\Psi_L
\right\}\,;\nonumber\\[2mm]
Q_L &=&\int dx \left\{\left(
\partial_L\bar \phi \right)\Psi_L + i  m\bar\phi\Psi_R
+i\frac{\bar w}{\bar\phi}\bar\Psi_R
\right\}\,,\nonumber\\[2mm]
\bar Q_L &=&\int dx \left\{\left(\partial_L\phi \right)\bar \Psi_L -
i \bar m\phi\bar \Psi_R -i\frac{ w}{\phi}\Psi_R
\right\}\,,
\label{fifteen}
\eeqn
where
\beq
\partial_R =
\partial_0- \partial_1\,,\qquad \partial_L = \partial_0 + \partial_1\,.
\label{rlder}
\eeq
Needless to say that
all supercharges are neutral with respect to $q_{\rm U(1)}$.

The following superalgebra ensues:
\beqn
\{\bar Q_L , Q_L\} &=& H+ P\,,\quad \{\bar Q_R , Q_R\} = H-P\,;
\label{sixteen}
\\[2mm]
\{Q_L , \bar  Q_R\} &=&m q_{\rm U(1)}- im \, \int dx\, \partial_x\, h\,,
\label{seventeen}
\\[2mm]
\{Q_R , \bar  Q_L\}&=& \bar m q_{\rm U(1)} + i \bar m \, \int dx\, \partial_x\, h
\,;
\label{eighteen}
\\[2mm]
 \{Q_R , Q_R\}&=&2 \bar m\,\bar w\int dx\,,\qquad
 \{\bar Q_R , \bar  Q_R\}=2 m\, w\int dx\,;
\label{nineteen}
\\[2mm]
\{Q_L , Q_L\}&=& -2  m\bar w\int dx \,,\qquad \{\bar Q_L , \bar  Q_L\}=-2
\bar m\, w \int dx \,;
\label{twenty}
\\[2mm]
\{Q_L , Q_R\}&=&-2 i \, \bar w\int dx\,\partial_x\, \ln\bar\phi\,,\qquad
\{\bar Q_L , \bar Q_R\}=  2 i \,  w\int dx\,\partial_x\, \ln \phi\,,
\label{twentyone}
\eeqn
where $(H,P)$ is the energy-momentum operator,
$$
(H,P) = \int dx \theta^{0\mu }\,,\qquad \mu =0,1\,,
$$
and $\theta^{\mu\nu}$ is the energy-momentum tensor.
Finally,   the function $h$ in Eqs. (\ref{seventeen}), (\ref{eighteen}) is  
\beq
h=\phi \bar{\phi}\,.
\label{twentytwo}
\eeq
 (The function $h$ is sometimes called
``the hamiltonian of the
isometry." Indeed, if we
formally consider the K\"ahler target space as a symplectic
manifold, i.e. as a phase space,
with the symplectic form $\Omega$ being the K\"ahler form,
$$
\Omega=-i d\phi \wedge d\bar{\phi}
$$
in our example,
and view $h$ as a ``hamiltonian,"
then the isometry transformations
 become a dynamical flow on the phase space,
$$
\partial_{\bar{j}}h = v^i \Omega_{i\bar{j}},\qquad
\partial_i h = -\bar{v}^{\bar{j}} \Omega_{i\bar{j}}\,.
$$
Warning: the function $h$ has nothing to do with the energy operator $H$ above.)

Equation (\ref{twentytwo}) is specific for  flat metric. Introduction
of a nontrivial K\"ahler metric   modifies
the expressions for supercharges, e.g. 
\beq
Q_R =\int dx \left\{G\left[ \left(
\partial_R\bar \phi \right)\Psi_R + i \bar m\bar\phi\Psi_L\right]
-i\frac{\bar w}{\bar\phi}\bar\Psi_L\right\} ,
\label{mod15}
\eeq
and so on. This entails, in turn, that the hamiltonian
of the isometry
$h$ must be modified appropriately too, 
see e.g. Sect.~\ref{XXX} where we treat the example of the $C^*$ model,
a ``relative''  of CP(1). 
 Equations (\ref{sixteen}) through (\ref{twentyone}) are valid in the
general case,
for {\em arbitrary} metric $G_{\bar j\, i}$, not only for the flat metric.
Equations (\ref{seventeen}) and (\ref{eighteen}) get extended by   anomalous terms,
 to be discussed in Sect.~\ref{anomr}.
A novel feature which we would like to stress is the
occurrence of central charges in Eqs. (\ref{nineteen}) and  (\ref{twenty})
that do not vanish if and only if both  twisted mass and   superpotential
are present. Note that they are proportional to the spatial size of the system.
Therefore, they are relevant to vacua rather than to BPS solitons,
as is the case for all other central charges.

\section{Implications for vacua}
\label{impliforvacua}
\setcounter{equation}{0}

In this section we will discuss implications of the algebra (\ref{sixteen})
--- (\ref{twentyone}).
For the time being we are interested in the structure of the
vacuum states rather than in solitons.
Therefore, we will switch off all {\em topological}
charges in  Eqs. (\ref{seventeen}), (\ref{eighteen}) and (\ref{twentyone}).
In addition we will set $q_{\rm U(1)}=0$. Our focus is the novel central charge
induced by combining twisted mass and superpotential,
Eqs. (\ref{nineteen}) and (\ref{twenty}).

For what follows it is convenient to introduce
two phases
\beq
e^{i\beta_R} = -\frac{|w/\bar m|}{w/\bar m}\,,\qquad
e^{ i\beta_L} = \frac{|w/ m|}{w/ m}\,.
\label{twentyfour}
\eeq
Suppose we are interested in the vacuum states, i.e.
terms with derivatives in the expressions for supercharges can be discarded.
It is not difficult to see that the following two supercharges
\beq
Q_3 = e^{- i\beta_R/2}Q_R + e^{i\beta_R/2}\, \bar Q_R \quad\mbox{and}\quad
Q_4 = e^{- i\beta_L/2} Q_L+ e^{i\beta_L/2}\,\bar Q_L
\label{twentyfive}
\eeq
annihilate the vacua, $Q_{3,4}|\rm vac\rangle =0$. Moreover,
\beq
Q_3^2=Q_4^2= H- |Z|\,,
\label{twentysix}
\eeq
where
\beq
 |Z| = 2 L |wm|\,,
\label{twentyseven}
\eeq
and $L$ is the spatial size of the system. 
The supercharges $Q_{3,4}$ (as well as $Q_{1,2}$ below) 
are defined in such a way that they are
Hermitean.

\begin{figure}[h]  
\epsfxsize=7cm
\centerline{\epsfbox{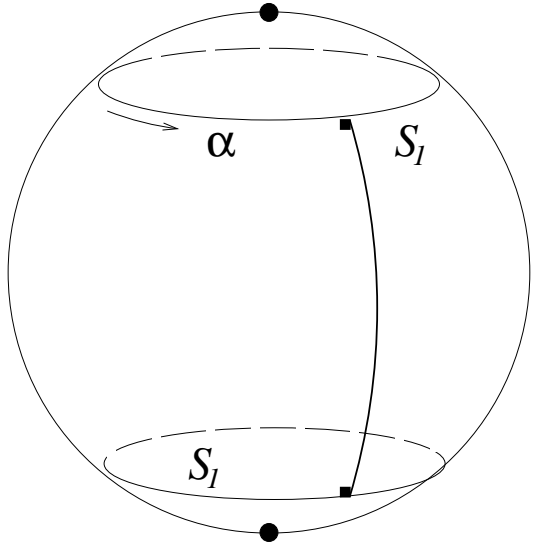}}
 \caption{The vacuum manifold in the simplest model
with twisted mass and superpotential, see (\ref{Aone}).
The target space is $S_2$ with the punctured  poles
(the poles are marked by closed circles).  The vacuum manifold consists
of two disconnected $S_1$ submanifolds. The trajectory running along a meridian
and interpolating between vacua belonging to the
first and second $S_1$ respectively 
(thick solid line) is a 1/4 BPS
saturated kink.}
\label{lstwmspfig} 
\end{figure}

As was repeatedly mentioned, in the general case
the only constraint on the metric is as follows:
the K\"ahler potential we deal with must be invariant
under the phase rotation of the fields $\phi,\,\,\bar\phi$, see Eq. (\ref{thirteen}).
A generic potential invariant under these rotations  
has the form
\beq
K(\phi^i\,,\,\,\overline{\phi^j}) = K(\phi^i\overline{\phi^j})\,.
\eeq
For instance, the K\"ahler potential
of the CP($N-1$) model,
$$
K_{{\rm CP}(N-1)} = \ln\left( 1 + \sum_{i=1}^{N-1} |\phi^i|^2
\right)\,,
$$
does the job. Needless to say that
the metric
$$
G_{i\bar j} =\frac{\partial }{\partial \phi^i }\, \frac{\partial }{\partial \bar\phi^{\, \bar j} }\,
K (\phi \bar\phi )
$$
is then also invariant. If we have a single field $\phi$
(and its complex-conjugated $\bar\phi$) 
$$
G_{\bar j\,i} \to  G(\bar\phi\phi )\,.
$$
where we assume  $G$   to be positive.
It is obtained from the K\"ahler potential $K(\phi \bar{\phi})$
as follows:
$$G(x)=K'(x)+x K''(x) \,,\qquad x = \phi \bar{\phi}.$$ 
Then the vacuum value of the field $\phi$ is
\beq
\phi_{\rm vac} = f_0\,\, e^{i\alpha}\,,
\label{twentyeight}
\eeq
where the parameter $f_0$ is any {\em positive} root of the equation
\beq
f^2 = \left| \frac{w}{m}\right| \, \frac{1}{G(f^2)}\, .
\label{twentynine}
\eeq
The phase
$\alpha$ is arbitrary. In other words, we find a continuous {\em compact}
vacuum manifold (see Fig.~\ref{lstwmspfig}). The number of positive 
roots of Eq.~(\ref{twentynine}) mod 2
is a topological invariant related to the index introduced in Sect. \ref{indices}.

Equation (\ref{twentynine}) is obtained by requiring
$Q_3$ (or $Q_4$) to annihilate vacuum. 
 The vacuum fields $\phi_{\rm vac},\,\,\bar\phi_{\rm vac}$
are coordinate-independent; therefore, all terms with derivatives
in Eq. (\ref{mod15}) and similar for $\bar Q_R$
can be dropped. The very same equation (\ref{twentynine})
is obtained through Bogomolny completion
of the potential term in the Hamiltonian,
$$
{\cal E} = G\left(\bar m\bar\phi +e^{i\beta_R}\,G^{-1}\, \frac{w}{\phi}\right)
\left(  m \phi +e^{-i\beta_R}\,G^{-1}\, \frac{\bar w}{\bar\phi}\right) + 2|wm|\,.
$$
 Equations (\ref{twentyeight}) and (\ref{twentynine})
assume that $P_{\rm vac}=0$.
The ground state energy density  ${\cal E}_{\rm vac}$ is
\beq
 {\cal E}_{\rm vac} = 2|mw|\neq 0\,.
\label{thirtythree}
\eeq

Normally this would signal spontaneous breaking of all four supercharges.
The well-known theorem reads that in extended supersymmetries one cannot break
spontaneously  a
part of it:
${\cal E}_{\rm vac}  \neq 0$ necessarily implies that all supercharges are broken.
However, in our case the
unconventional  central extension of   superalgebra,
 Eqs. (\ref{nineteen}) and (\ref{twenty}), invalidates this theorem.
Thus, in the sigma model with twisted mass {\em and}
superpotential, ${\cal N}=2$ symmetry is {\em spontaneosly} broken down to
${\cal N}=1$ provided
that Eq. (\ref{twentynine}) has at least one positive root.
(It can have more than one root, though.)

Now, it is
instructive to consider
two remaining combinations of the supercharges --- those orthogonal
to $Q_{3,4}$,
\beq
Q_1 = i\left( e^{- i\beta_R/2}Q_R - e^{i\beta_R/2}\,
\bar Q_R\right) \quad\mbox{and}\quad
Q_2 =i\left(  e^{- i\beta_L/2} Q_L - e^{i\beta_L/2}\,\bar Q_L\right).
\label{thirty}
\eeq
They do {\em not} annihilate the vacuum state since
\beq
Q_1^2=Q_2^2= H+  |Z|\,.
\label{thirtyone}
\eeq
Rather, acting on   vacuum, they produce
Goldstino,
\beq
\psi_{\rm G } = \frac{1}{\sqrt 2}\,
\left(\begin{array}{c}
e^{-i\gamma_R}\Psi_R +e^{i\gamma_R}\bar\Psi_R\\[3mm]
e^{-i\gamma_L}\Psi_L +e^{i\gamma_L}\bar\Psi_L
\end{array}
\right),
\label{goldstino}
\eeq
where\,\footnote{Note that $\beta_R +2\mbox{arg}\,(m)$ is related to the phase of
$w\bar m$ while $\beta_L  -2\mbox{arg}\,(m)$ to the phase of $w m$.} 
\beq
\gamma_R=\frac{\beta_L}{2}+\alpha -\mbox{arg}\,(m)\,,\qquad
\gamma_L=\frac{\beta_R}{2}+\alpha +\mbox{arg}\,(m)\,.
\label{goldstinoph}
\eeq
The Goldstino field is a Majorana field which pairs up with
the massless scalar corresponding to the would-be Goldstone 
boson \footnote{Global symmetries such as U(1) in
two dimensions cannot be spontaneously broken~\cite{Coleman:ci}. Quantum
fluctuations ``smear" the vacuum wave function
  over the vacuum manifold. Nevertheless, in the case of
U(1), a massless boson persists in the spectrum. Its coupling to all 
U(1) noninvariant operators vanishes, however. It is derivatively coupled to the
U(1) current.}
 of U(1)
to form a massless ${\cal N}=1$ supermultiplet. (The above massless scalar
is in fact the phase of $\phi$.)
The second Majorana spinor field, orthogonal
to (\ref{goldstino}),
\beq
\psi_{m} = \frac{i}{\sqrt 2}\,
\left(\begin{array}{c}
e^{-i\gamma_R}\Psi_R -e^{i\gamma_R}\bar\Psi_R\\[3mm]
e^{-i\gamma_L}\Psi_L -e^{i\gamma_L}\bar\Psi_L
\end{array}
\right),
\label{massivemaj}
\eeq
 pairs up with the remaining massive scalar field to form the second 
${\cal N}=1$ supermultiplet, with mass $\neq 0$.
If we parametrize  
$\phi = f_0 + 2^{-1/2} (\sigma + i\rho)$,
then the quadratic part of the Lagrangian
takes the form
\beqn
{\cal L}^{(2)} &=& \frac{1}{2}\, G(f_0^2)\left(\partial_\mu\sigma\, 
\partial^\mu\sigma -\tilde{m}^2\sigma^2 + 
\bar\psi_{m }\not\!\partial\, \psi_{m } -
\tilde{m}\bar\psi_{m }\psi_{m }
\right.
\nonumber\\[3mm]
&+&\left.
\partial_\mu\rho\, \partial^\mu\rho+ \bar\psi_{\rm G }  \not\!\partial\, 
\psi_{\rm G }
\right)\,,
\label{lquad}
\eeqn
where
\beq
\tilde{m} = |m|\,\left|
2+f\frac{\partial}{\partial f}\,\ln\, G
\right|_{f=f_0}\,.
\eeq
Let us parenthetically note that
in the model (\ref{Aone})
\beq
\tilde{m} =2|m|\,\frac{\sqrt{\beta^2-1}}{\beta}\,,
\label{divo}
\eeq
where the parameter $\beta$ is defined in Eq. (\ref{Aeleven}).

If Eq. (\ref{twentynine}) has no positive roots,
all four supercharges act on the vacuum nontrivially. Then
${\cal N}=2$ supersymmetry is completely spontaneously broken.

\section{Counting vacua and kinks,  or what replaces 
the CFIV index}
\label{indices}
\setcounter{equation}{0}

As was explained above,
the fermion charge is not defined  in the model at hand.
Of course,
the fermion parity $(-1)^F$
remains well-defined (the latter
is in contradistinction to what happens \cite{LSV} in ${\cal N}=1$ models).

The absence of the fermion number precludes us from using the   CFIV
index.\footnote{The Witten index always vanishes here.}
However,
since ${\cal N}=2$ superalgebra remains valid
(albeit centrally extended), there should exist
an index replacing the CFIV index. The replacement must
be formulated exclusively in internal algebraic terms,
as was done in Ref. \cite{LSV} that treated
${\cal N}=1$ two-dimensional models where
neither $F$ nor $(-1)^F$ was   defined for BPS states.

Here we   suggest indices for counting 1/2
BPS-saturated states (vacua)  and 1/4 BPS-saturated  states
(kinks)
in the problem at hand (centrally
extended ${\cal N}=2$). Our construction 
is a direct generalization of that of Ref. \cite{LSV}.

\subsection{A brief  reminder of 
the ${\cal N}=1$ story}
\label{brief}

In the ${\cal N}=1$ case we study 
the representations of the superalgebra  
\beq
Q_1^2= H+Z \,,\qquad  Q_2^2=H-Z\, ,
\eeq
where $Z$ is assumed to be  positive.

For non-BPS states, $H>Z$, the representation is obviously two-dimensional. Each supermultiplet is a doublet.
The BPS states correspond to one-dimensional representations
of the algebra 
which can be of two types, $v_{\pm}$, such that
\beq
Q_2\, v_{\pm}=0\,,\qquad  Q_1 \, v_{\pm}=\pm \sqrt{2Z} \, v_{\pm}\,.
\label{twot}
\eeq
The index is the difference between the numbers of the BPS 
states with the positive
and negative eigenvalues of $Q_1$.
Note, that if the fermion parity $(-1)^F$ is a symmetry
of the theory, then the above  index necessarily
vanishes.
By fermion  parity we mean   an operator such that
\beq
(-1)^F \, Q_i \,  (-1)^F = - Q_i\,,\qquad i=1,2\,.
\eeq
Indeed, assume we have a state $v_{+}$.
Then the state $(-1)^F  v_+$  is in fact $v_-$,
\beq
Q_1\left\{ (-1)^F \, v_{+}\right\} =(-1)^F\left(-Q_1\right) v_{+}
= -\sqrt{2Z} \left\{ (-1)^F \, v_{+}\right\}\,.
\label{vpvm}
\eeq
In other words, the existence of $(-1)^F$ implies that
the states appear in pairs $v_{+},\,\, v_-$.

\vspace{1mm}

For a generic representation (which can
be  reducible)   the proper index can be introduced as follows:
\beq
{\cal I}_1= {\rm Tr }\, \frac{Q_1}{\sqrt{2Z}} \exp{(-t \, Q_2^2)}\,.
\label{lsvold}
\eeq
From this expression it is clear that the index 
${\cal I}_1$, being an integer
and a smooth function of the parameters of the system, is constant
on the space of parameters of the system.
For short (one-dimensional) multiplets
the value of ${\cal I}_1$ is $\pm 1$. For long multiplets
${\cal I}_1=0$. 

\subsection{Counting   1/2 BPS states in ${\cal N}=2$ SUSY}
\label{chalf}

The 1/2 BPS-saturated states in ${\cal N}=2$ SUSY appear when   superalgebra  takes the form
\beq
Q_1^2=Q_2^2=H+|Z|\, ,\qquad Q_3^2=Q_4^2=H-|Z|\, ,
\eeq
(cf. Eqs. (\ref{twentysix}) and (\ref{thirtyone})).
This algebra has four-dimensional irreducible representations
--- these are long non-BPS 
representations with  $H>|Z|$.
If $H=|Z|$ the irreducible representations are short, 
  two-dimensional supermultiplets of the BPS-saturated states.

\vspace{1mm}

Cecotti {\em et al.} proposed 
\cite{CFIV} an index,
\beq
I_{\rm CFIV} ={\rm Tr}\,  F (-1)^F e^{-t(H-|Z|)}\,,
\label{cvifdefi}
\eeq
 which is 
integer-valued, saturated by short multiplets,
 and   smoothly depends on parameters of the system ---
thus, it is constant on the space of parameters\,\footnote{
In fact, the definition of the CFIV index in the original papers
\cite{CFIV} is somewhat different from that given in Eq.~(\ref{cvifdefi}).
Cecotti {\em et al.} proved that their index
does not change under continuous variations of $D$ terms. }.
However, it requires the existence of a conserved
fermion  number   $F$ with the appropriate
commutation relations.
In our theory we do not have this luxury.  In the case under consideration we
do have
 the conserved fermion parity
$(-1)^F$. Therefore, we suggest to   replace $I_{\rm CFIV}$
by a new index ${\cal I}_{2,1/2}$ defined as follows:
\beq
{\cal I}_{2,1/2} = \frac{1}{2}\, {\rm Tr} \,  (-1)^F \, \frac{i\, Q_1 Q_2}{2|Z|} \, e^{-t(H-|Z|)}\,.
\label{new2half}
\eeq

The meaning of the above index is easy to assess. 
Again, as in Sect. \ref{brief}, there are two types of BPS states, $v_\pm$
(now both $v_+$ and $v_-$ are in fact doublets, say, $v_{+,\alpha}$, where $\alpha = 1,2$;
we suppress the $\alpha$ index for brevity). The states $v_\pm$ 
can be defined through the relation\,\footnote{One should proceed with care 
here since there is no mass gap in the theory.
The supercharges $Q_{1,2}$ correspond to the
spontaneously broken supersymmetries;
their action creates Goldstinos.
It is necessary to regularize the theory in
the infrared, say, by putting it in a large spatial box.}
\beq
  i\, Q_1 Q_2  \, v_{\pm} = \pm \, 2|Z| \, (-1)^F \, v_{\pm}\,. 
\label{thrrel}
\eeq
Then, the index (\ref{new2half}) obviously 
counts the difference between the
number of the BPS doublets $v_{+}$ and that of
  $v_{-}$'s. Long multiplets necessarily consist of
pairs
$\{ v_+\,,\,\, v_-\}$; hence for long multiplets ${\cal I}_{2,1/2}$ vanishes.

\subsection{Counting   1/4 BPS states in ${\cal N}=2$ SUSY}
\label{cquarter}

1/4 BPS saturated states in the model under
consideration are kinks, to be discussed in  Sect.
\ref{seckinks}. 
There are four distinct central charges in the corresponding
superalgebra presented in Eqs. (\ref{1/4one}) through (\ref{1/4four}).
In the 1/4 BPS kink sector we loose   $(-1)^F$.
If the theory is regularized in the infrared, see the previous
footnote, one can say that short 1/4 BPS-saturated multiplets are doublets,
while long multiplets are quadruplets.
This follows from the algebra (\ref{1/4one}) through (\ref{1/4four}),
with $Q_4=0$ on the BPS states, and all four supercharges 
operative for non-BPS states.

From the algebraic standpoint the fact that the short kink 
multiplets
are two-dimensional is perfectly obvious. 
However, it is instructive to ask a physical question:
``what is the multiplicity of  the kink particles?" The point is that,
as has been just mentioned in Sect. \ref{chalf},
 double degeneracy is the intrinsic  property
of the 1/2 BPS-saturated vacua. It is this double degeneracy
that  is
inherited by the 1/4 BPS  kink multiplets.
If we consider kinks interpolating between specific given vacua,
then they are unique. Therefore, kinks considered as particles, 
form one-dimensional representations of superalgebra,
see Sect.~\ref{seckinks}.

Returning to superalgebra
(\ref{1/4one}) through (\ref{1/4four}), one can define two types of the BPS states
as follows:
\beq
i\, Q_1\, Q_2\, Q_3\,  v_\pm =\pm\sqrt{(Z_4+Z_1)(Z_4+Z_2)(Z_4-Z_3)}\,v_\pm\,.
\eeq
Then 
\beq
{\cal I}_{2,1/4} = \frac{1}{2}\, {\rm Tr} \,  
  \frac{i\, Q_1 Q_2 Q_3}{\sqrt{(Z_4+Z_1)(Z_4+Z_2)(Z_4-Z_3)}} \, \, e^{-tQ_4^2}
\label{new2quarter}
\eeq
is the desirable index counting the difference between the
numbers of $v_+$ and $v_-$. 

Note, that the index ${\cal I}_{2,1/4}$, just like
its ${\cal N}=1$ counterpart (see Eq. (\ref{lsvold})),   
can be nonvanishing only if the fermion parity is broken.
Indeed, if the 
 fermion parity is defined,
$
(-1)^F Q_i (-1)^F = - Q_i\,\quad i=1,...,4,
$
then  
$v_+$ and  $v_-$ appear in pairs $\{v_-,\,\,v_+\}$,
$$
v_-=(-1)^F v_{+}\,.
$$

\section{Showcase example of 
$C^*$ model (descendant of CP(1))}
\label{XXX}
\setcounter{equation}{0}

Here we briefly illustrate our construction
starting from 
 the simplest and the most popular ${\cal N}=2$ sigma model, CP(1).
We deform this model by adding twisted mass and superpotential,
following the strategy outlined in Sect. \ref{twistedmass}. 
The CP(1) model is the sigma model on the two-dimensional sphere.
It is described by one complex field $\phi$ and one
complex two-component spinor $\Psi$. The corresponding
Lagrangian is obtained from Eq.~(\ref{one})
by dropping the indices $i$ and $\bar j$. The target space of the deformed
model has the topology of cylinder; the model is referred to as 
the $C^*$ model.

The Lagrangian  including twisted mass and   superpotential
can be written as
\beqn
{\cal L}_{\,  C^*}&=& G\, \left\{
\partial_\mu\bar\phi\partial^\mu\phi -|m|^2\, {\phi\bar\phi}
\right.
\nonumber\\[3mm]
&+&\frac{i}{2}\left(\bar\Psi_L\stackrel{\leftrightarrow}{\partial_R}\Psi_L + 
\bar\Psi_R\stackrel{\leftrightarrow}{\partial_L}\Psi_R
\right)-\frac{1-\bar\phi\phi}{\chi} \left(m\bar\Psi_L\Psi_R +\bar m \bar\Psi_R\Psi_L
\right)
\nonumber\\[3mm]
&-&\frac{i}{\chi}\, \left. \left[\bar\Psi_L \Psi_L
\left(\bar\phi\stackrel{\leftrightarrow}{\partial_R}\phi
\right)+ \bar\Psi_R \Psi_R
\left(\bar\phi\stackrel{\leftrightarrow}{\partial_L}\phi
\right)
\right]
-
\frac{2}{\chi^2}\bar\Psi_L\Psi_L \bar\Psi_R\Psi_R
\right\}
\nonumber\\[3mm]
&-& G^{-1} |w|^2\,\frac{1}{\bar\phi\phi} +\left\{\frac{1-\bar\phi\phi}{\chi}\,\, 
\frac{w}{\phi^2}\, \Psi_R\Psi_L + \mbox{h.c.}\right\}
\nonumber\\[3mm]
&+&
\frac{i \theta}{2\pi}\,\frac{1}{\chi^2}\,
\varepsilon^{\mu\nu} \partial_\mu\bar\phi
\partial_\nu \phi \,,
\label{Aone}
\eeqn
where the metric $G$
\beqn
G&\equiv& \frac{2}{g^2}\,{\cal G} = \frac{2}{g^2}\,\frac{1}{(1+\phi\bar\phi)^2}\,, 
\nonumber\\[3mm]
{\cal G} &=&\frac{1}{\chi^2}\,,\quad
\chi \equiv 1+\phi\bar\phi\,.
\label{Atwo}
\eeqn
The target space is $S_2$.
Note that
the only nonvanishing Christoffel symbols
are
\beq
\Gamma = - 2\, \frac{\bar\phi}{\chi}\,,\qquad \bar\Gamma =
- 2\, \frac{ \phi}{\chi}\,,
\label{Athree}
\eeq
while the Ricci tensor in the case at hand is related to the metric as follows
\beq
R = 2\,{\cal G}= \frac{2}{\chi^2}\,.
\label{Afour}
\eeq
The scalar curvature ${\cal R}$ is the inverse radius
squared of
the target space sphere,
$
{\cal R} =  {g^2}\,.
$
The vector fields on the CP(1) target
space are
\beq
v=\phi, \qquad \bar v= - \bar \phi\,.
\label{Asix}
\eeq
Thanks to the existence of this
isometry one can introduce twisted mass
which
  breaks O(3) of the undeformed model down to  U(1).
Besides twisted mass we add a superpotential
  term preserving U(1) and
${\cal N}=2$ supersymmetry; it
is given in the last but one line of Eq. (\ref{Aone}).
The last line is the $\theta$ term.
The conserved
supercurrent has the form
\beq
J^\mu =
\left( \partial_\lambda\bar\phi \right)\gamma^\lambda\gamma^\mu \Psi
-i\, \bar m\, \bar\phi \,\gamma^\mu\,  \Psi
+ i \frac{\bar w}{\bar \phi}\,
\gamma^\mu\Psi^\dagger\,.
\label{Aseven}
\eeq
Finally, we have to specify the ``hamiltonian"
$h$ in Eqs. (\ref{seventeen}) and (\ref{eighteen}),
\beq
h= \frac{2}{g^2} \, \frac{\bar\phi\phi}{1+\bar\phi\phi}\,.
\label{Aeight}
\eeq
Alternatively the ``hamiltonian" can be rewritten as
$h = - (2/g^2)(1+\bar\phi \phi )^{-1}$.
The vacuum manifold consists of two disconnected $S_1$'s (Fig.~\ref{lstwmspfig}),
\beq
\phi_{\rm vac} = \left(\beta \pm \sqrt{\beta^2-1}\right) e^{i\alpha}
\label{Aten}
\eeq
where we introduce a parameter $\beta$,
\beq
\beta =\sqrt{\left|\frac{m}{w}
\right|\, \frac{1}{2g^2}}\,,
\label{Aeleven}
\eeq
and it is assumed that $\beta >1$. Otherwise
${\cal N} =2$ SUSY is completely broken. Elementary excitations
are: (i) a massleess ${\cal N} = 1$ supermultiplet,
one Goldstone boson plus one Goldstino;
(ii)
a massive ${\cal N} = 1$ supermultiplet, with mass
\beq
\tilde{m} =2|m|\,\frac{\sqrt{\beta^2-1}}{\beta}\,.
\label{masplit}
\eeq

The fact that there are two solutions for $\phi_{\rm vac}$ at 
$\beta > 1$ and no solutions at $\beta <1$
means that the index ${\cal I}_{2,1/2}=0$
in the model at hand. The situation when the classical BPS vacua
exist while  the   index is vanishing is always a special case.
At the quantum level BPS vacua may or may not exist.
Let us show that they do exist in the field theory
under consideration. Of crucial importance is the fact that
this theory is weakly coupled.

Indeed, assume that the classical vacua at $ {\cal E}_{\rm vac} = 2|mw|$
are lifted.
Then all four supercharges act nontrivially on the
vacuum state, and all four supersymmetries are
spontaneously broken. We should have then 
two Majorana Goldstino fields rather than one.
The fermion mass matrix in the classical vacua splits in two blocks ---
one with the vanishing eigenvalue and another with the eigenvalue
$\tilde{m} =2|m|\, {\sqrt{\beta^2-1}}/{\beta}$.
If $\beta$ is not close to unity, quantum corrections at weak coupling
cannot move $\tilde{m}$ to zero. 
There is no candidate field for the second Goldstino field.
Hence, the assumption above is wrong, and the vacua
  at $ {\cal E}_{\rm vac} = 2|mw|$
are not lifted.

As $\beta$ tends to unity, two vacuum manifolds (circles)
 in Fig.~\ref{lstwmspfig}
coalesce. At $\beta =1$ an extra massless fermion field develops
which becomes the second Goldstino field.
At $\beta >1$ all supersymmetry is spontanesouly broken.

Another interesting limit is $\beta \to \infty$.
In this limit we switch the superpotential off, and 
the vacuum manifolds degenerate into two points,
the north and the south poles.

We will consider kinks in the $C^*$ model momentarily.
For generic values of $\beta$ from the interval
$1<\beta <\infty$ these kinks have a single bosonic modulus, 
the kink center, and are 1/4 BPS-saturated. 
In the limit  $\beta \to \infty$ the superpotential drops out,
the angle $\alpha$ (see Eq. (\ref{pasafive})) becomes a modulus,
and we recover 1/2 BPS-saturated kinks of the model treated in Ref.
\cite{Dorey}.

\section{Kinks}
\label{seckinks}
\setcounter{equation}{0}

Let us assume that Eq. (\ref{twentynine})
has more than one solution.
An example with two solutions has been just 
 considered in Sect.~\ref{XXX}.
Then the vacuum manifold consists of several disconnected submanifolds 
(Fig.~\ref{lstwmspfig}).
In this case one can (and should) include in the analysis
field configurations interpolating between a given vacuum from the first 
submanifold and its counterpart from the
second submanifold. Such interpolating trajectories are kinks.

From the purely algebraic standpoint,
${\cal N}=2$ superalgebra in two dimensions
admits a combination of  central charges corresponding to
1/4 BPS-saturated solitons\,\footnote{This fact was emphasized by A. Vainshtein
prior to the development of the model with twisted mass and superpotential
reported in  the present paper. We are grateful to him for
thorough discussions of 
(super)algebraic aspects. Recently  we have considered
the issue of BPS-saturated solitons in ${\cal N}=2$ models in  two dimensions
in a related context in Ref.~\cite{hou}. Versions of some  equations presented 
here
first appeared in this work. }.
To the best of our knowledge, none of two-dimensional models discussed in the 
literature exhibits such solitons, however. In fact, ${\cal N}=2$
sigma models with twisted mass {\em and} superpotential,
which we investigate here, seem to be the first example
where 1/4 BPS-saturated kinks with mass
\beq
M_{\rm kink} =\left|
{|m|}\, h -|w|\,\ln \,\bar\phi \phi
\right|^{\phi_{{\rm vac}2}}_{\phi_{{\rm vac}1}}
\label{pasaone}
\eeq
appear in a natural way. The definitions of $h$ and $\beta$
in the $C^*$ model are given in Sect.~\ref{XXX}. In this particular
model the right-hand side of Eq.~(\ref{pasaone})
reduces to
\beq
M_{\rm kink} =\frac{2|m|}{g^2}
\left(\frac{\sqrt{\beta^2-1}}{\beta}-\frac{1}{2\beta^2}\,
\ln\frac{\beta + \sqrt{\beta^2-1}}{\beta - \sqrt{\beta^2-1}}
\right)\,.
\label{pasaonep}
\eeq
The point  $\beta =1$ is singular, in full accordance with the
discussion in Sect.~\ref{XXX}.

The 1/4 BPS-saturated kinks exist
in generic models with twisted mass, 
superpotential, and more than one solution 
of Eq.~(\ref{twentynine}). To study them in more detail
we turn to the model (\ref{Aone}). 
In this model, for static field configurations, we can
represent the energy functional in the form
\beqn
{\cal H} &= &
G
\left(-\partial_x\bar\phi +|m|\bar\phi - G^{-1}\,\frac{|w|}{\phi}
\right)
\,
\left(-\partial_x\phi +|m|\phi - G^{-1}\,\frac{|w|}{\bar\phi}
\right)
\nonumber\\[3mm]
&+&\int_{-\infty}^{\infty}\, dx\, 
\left\{
2|mw| -|w|\,\partial_x\,\ln \left(\bar\phi\phi\right)
+{|m|}\, \partial_x\, h
\right\}\,.
\label{pasatwo}
\eeqn
The expression in the first line vanishes
provided that
\beq
\partial_x \phi = |m|\phi - G^{-1}\,\frac{|w|}{\bar\phi}\,.
\label{pasathree}
\eeq
If Eq. (\ref{pasathree}) has a solution with
the boundary conditions
\beq
\phi\to \phi_{{\rm vac}1}\,\,\,\mbox{at}\,\,\, x\to -\infty ,\,\,\,
\mbox{and}\,\,\,
\phi\to \phi_{{\rm vac}2}\,\,\,\mbox{at}\,\,\, x\to \infty\,,
\label{pasafour}
\eeq
then the Bogomolny bound will be achieved, and the solution at hand
will be a 1/4 BPS-saturated kink.
We will first give the solution and then
present a supercharge which
is conserved on this solution.

The solution of Eq. (\ref{pasathree})
is 
\beq
\phi (x) =e^{i\alpha}\,
\sqrt{\frac{|\phi_{{\rm vac}1}|^2+|\phi_{{\rm vac}2}|^2\, 
e^{\tilde{m}x}}{1+e^{\tilde{m}x}}}
\label{pasafive}
\eeq
where $\alpha$ is an arbitrary phase
marking the initial and final vacua, and
$\tilde m$ is defined in Eq. (\ref{divo}).  (Warning: $\alpha$  is no
modulus, since the corresponding zero mode is non-normalizable, see Fig.~\ref{lstwmspfig}).
For each given ``initial" vacuum from the first submanifold
we have one bosonic kink; the ``final" vacuum is fixed.

It is instructive to have a closer look at superalgebra relevant
to the kink problem. Not to overburden the paper with 
combersome formulae we will not consider arbitrary phases
of the central charges, although this is certainly doable.
Instead, we will specify the values of parameters
in such a way that all central charges in Eqs. (\ref{sixteen})
through (\ref{twentyone}) are real and positive. 
To this end we set $\phi$ real ($\alpha =0$) and replace
\beq
m\to i m,\quad \bar m \to -im,\quad w\to -i w,\quad \bar w\to i w
\eeq
with real and positive $m$ and $w$ after the substitution.
Then
\beq
Q_1^2 = H+Z_1,\qquad Z_1 =2mw L- m\Delta h -w\Delta \ln\bar\phi\phi\,,
\label{1/4one}
\eeq
\beq
Q_2^2 = H+Z_2,\qquad Z_2 =2mw L+m\Delta h +w\Delta \ln\bar\phi\phi\,,
\label{1/4two}
\eeq
\beq
Q_3^2 = H-Z_3,\qquad Z_3 =2mw L- m\Delta h + w\Delta \ln\bar\phi\phi\,,
\label{1/4three}
\eeq
\beq
Q_4^2 = H-Z_4,\qquad Z_4 =2mw L+m\Delta h - w\Delta \ln\bar\phi\phi\,,
\label{1/4four}
\eeq
where the $L$ is the spatial size of system,
while the supercharges $Q_{1,...,4}$ are
the following linear combinations of the original ones:
\beq
Q_1 = \frac{1}{2 }\, \left(Q_R-Q_L+\bar Q_R-\bar Q_L
\right)\,,
\label{1/4onea}
\eeq
\beq
Q_2 = \frac{1}{2 }\, \left(Q_R+Q_L+\bar Q_R+\bar Q_L
\right)\,,
\label{1/4twoa}
\eeq
\beq
Q_3 = \frac{1}{2 i}\, \left(Q_R+Q_L-\bar Q_R-\bar Q_L
\right)\,,
\label{1/4threea}
\eeq
\beq
Q_4 = \frac{1}{2 i}\, \left(Q_R-Q_L-\bar Q_R+\bar Q_L
\right)\,,
\label{1/4foura}
\eeq
They are Hermitean. The momentum $P$ is set to zero (rest frame).
All anticommutators of the type
$\{Q_i , Q_j\}$ with $i\neq j$ vanish. The kink is annihilated by $Q_4$,
i.e. $H=Z_4$.
Then
\beq
Q_1^2=4mwL-2w\Delta \ln\bar\phi\phi ,\quad Q_2^2 = 4mwL+2 m\Delta h,\quad
Q_3^2 = 2 m\Delta h- 2w\Delta \ln\bar\phi\phi \,.
\eeq
The first two supercharges scale as $\sqrt L$ with $L$.
They correspond to those supersymmetries that
are spontaneously broken in the vacuum. Their action on the kink creates Goldstinos,
rather than normalizable fermion zero mode.
The action of $Q_3$ on the bosonic kink solution yields
a normalizable zero mode. 

Note that only one normalizable
 fermion zero mode is associated with the 
kink solution. Therefore, in the kink sector the fermion parity is 
lost, as is explained in detail in Ref. \cite{LSV}.
The kink supermultiplet is {\em one}-dimensional,
see also remarks in Sect.~\ref{cquarter}.

Changing $\alpha$ corresponds to   motion along the vacuum manifold,
(i.e. in the direction in the functional space ``perpendicular"
to the kink trajectory), i.e.
to excitation of the Goldstone quanta. It would be interesting to explore
kink interactions with these Goldstone quanta.
Furthermore,
our treatment of the 1/4 BPS-saturated kink was quasiclassical. What's to 
be done in the
future is to find out
what aspects will survive inclusion of nonperturbative effects along the 
lines of Ref. \cite{Dorey}.

\section{Loop corrections}
\label{cor}
\setcounter{equation}{0}

The central charges in Eqs. (\ref{nineteen}) and (\ref{twenty})
are 
holomorphic; they 
are not renormalized in loops, and so is the vacuum energy density
(\ref{thirtythree}). This follows from the fact that the
vacuum energy density is protected from loop corrections 
in the effective ${\cal N}=1$ low-energy theory by
${\cal N}=1$ supersymmetry which remains realized linearly.
This is in full accord with nonrenormalization
of the twisted mass parameter $m$ and the superpotential parameter $w$.

At the same time the central charges in Eqs. (\ref{seventeen}) and
(\ref{eighteen}) are renormalized in loops. Unlike the
$w=0$ case, 
holomorphy is ruined from
the very beginning by the condition 
(\ref{twentynine}), and 
  loop corrections go beyond
the one-loop order. Besides, these central charges 
acquire a quantum anomaly,  
which we will discuss momentarily. 

\section{  Anomaly}
\label{anomr}
\setcounter{equation}{0}

It is a
folklore statement that ${\cal N}=2$ sigma models in two dimensions present
a close parallel to ${\cal N}=1$ Yang-Mills theory in four dimensions
(for a pedagogical discussion of this issue see e.g. Ref. \cite{NPRep}).
The anomaly aspect is no exception. Superalgebra in
 ${\cal N}=1$ Yang-Mills theory   possesses a (tensorial)
central charge which emerges \cite{DSano} due to anomaly
generalizing the supermultiplet of geometric anomalies
in  $\partial_\mu a^\mu$, $\gamma^\mu J_{\mu }$ and $\theta^\mu_{\, \mu}$.
(Here $a^\mu$ is
supersymmetric gluodynamics axial current while $J_{\mu }$ is supersurrent.)
A similar anomalous central charge occurs in
  ${\cal N}=2$ sigma models
in two dimensions in the anticommutators $\{ Q_L \,,\,\bar  Q_R\}$
and $\{Q_R \,,\,  \bar  Q_L\}$.
We will present here the corresponding result for the undeformed model
(i.e. in the
limit $m\to 0$, $w\to 0$). In this limit, at the classical level these two
anticommutators vanish. Taking account of the anomaly induces
(anomalous) central charges.

To obtain these central charges one can exploit a
variety of  methods similar to those exploited  in Ref.
\cite{DSano}. Deferring a more detailed discussion
till a later publication,
we sketch here a derivation in broad touches.
Our derivation will be applicable to
compact homogenious symmetric
K\"{a}hler target spaces for which the K\"{a}hler metric $G_{\bar j\, i}$
is characterized by a {\em single coupling constant} $g^2$ which is usually
introduced as
\beq
G_{\bar j\, i} \equiv \frac{2}{g^2}\, {\cal G}_{\bar j\, i} \,.
\label{thirtyseven}
\eeq

There are four popular  classes of compact homogenious symmetric
K\"{a}hler target spaces;
 they are listed in Table 1.  The first class
is usually referred to as Grassmannian sigma models, of which
CP($n$) is a particular case.  CP($n$) models are obtained by setting $m=1$.
If $m=n=1$ we get the showcase CP(1) model.
An important feature crucial for our derivation of the anomaly is as follows: 
 for any symmetric K\"ahler space
the Ricci tensor $R_{i\bar j}$ is proportional to the metric \cite{KN}, namely,
\beq
R_{i\bar j} = b \, {\cal G}_{\bar j\, i}\,,
\label{thirtyeight}
\eeq
where
  $b$ stands for the coefficient
in  the   corresponding  Gell-Mann--Low function, see Table 1.

\begin{table}
\begin{center}
\begin{tabular}{|c|c|c|c|}\hline \hline
&
&
&
\\[-1mm]
${\cal T}_1=
\frac{{\rm SU}(n+m)}{{\rm SU}(n)\otimes{\rm SU}(m)\otimes{\rm U}(1)} $ &
 ${\cal T}_2= \frac{{\rm Sp}(n)}{{\rm SU}(n)\otimes{\rm U}(1)}$ &
${\cal T}_3= \frac{{\rm SO}(2n)}{{\rm SU}(n)\otimes{\rm SO}(2)}$ &
${\cal T}_4= \frac{{\rm SO}(n+2)}{{\rm SO}(n)\otimes{\rm SO}(2)}$
\\[3mm]
\hline
$b=m+n$ &
$ b= n+1$  &
 $ b= n-1$  &
$b= n$
\\
\hline\hline
\end{tabular}
\caption{Coefficients $b$ in the
Gell-Mann--Low function for four classes of
${\cal N}=2$ compact homogenious symmetric
K\"{a}hler sigma models.}
\end{center}
\end{table}

Now we are ready 
to outline derivation of the anomaly.
 The  conserved supercurrent of the model at hand 
has the form
\beq
J^\mu = G_{\bar j\, i} \left(\partial_\lambda \bar\phi^{\, \bar j}
\right)\gamma^\lambda\gamma^\mu \, \Psi ^i +...
\label{thirtyfivep}
\eeq
Naively $\gamma_\mu J^\mu$  vanishes,
which coresponds to the (super)conformal invariance
at the classical level. As well-known, the (super)conformal invariance
is anomalous --- it is broken at the quantum level.
The simplest way to see this is to calculate 
$\gamma_\mu J^\mu$ in $D=2-\varepsilon$ dimensions, rather than in two
dimensions. Then
\beq
\gamma_\mu J^\mu = \varepsilon\,  G_{\bar j\, i} \left(\partial_\lambda \bar\phi^{\, \bar j}
\right)\gamma^\lambda \, \Psi ^i\,.
\label{epsilonreg}
\eeq
Although formally the right-hand side vanishes at $\varepsilon\to 0$,
this overall factor $\varepsilon$ is cancelled by $1/\varepsilon$
residing in the coupling constant,
$$
\frac{1}{g^2} = \frac{1}{g^2_{\rm ren}} + \frac{b}{4\pi}\, \frac{1}{\varepsilon}\,.
$$
Taking into account Eqs. (\ref{thirtyseven}) and (\ref{thirtyeight})
we thus obtain
the anomaly in $\gamma_\mu J^\mu$,  
\beq
\gamma_\mu J^\mu = \frac{R_{i\bar j}}{2\pi}\,
\left(\partial_\lambda \bar\phi^{\, \bar j} \right)\gamma^\lambda \Psi^i
\,.
\label{thirtysix}
\eeq
[Let us parenthetically note that by the same token
one can readily obtain
 the anomaly in the trace of the energy-momentum tensor,
\beq
\theta^\mu_\mu = \frac{R_{i\bar j}}{2\pi}\,\left(
 \partial^\mu\bar\phi^{\,\bar j}\, \partial_\mu\phi^i
+\frac{i}{2} \,
\bar\Psi^{\bar j} \!\stackrel{\leftrightarrow}{\not\!  \partial}\Psi^{i}
+...\right)\,,
\label{anoenmo}
\eeq
where the ellipses denote terms of higher order in
$\partial\phi, \,\, \Psi$.]

The next step is to show that the anomaly (\ref{thirtysix})
implies the emergence of the anomalous central charges.
Indeed, the general representation for the anticommutator
$\{\bar Q , J_\mu\}$ compatible with
superalgebra and the supercurrent conservation is as follows:
\beqn
\left\{\bar Q _\beta, J_{\mu\,,\alpha}\right\} &=&
\left(\gamma^\nu\right)_{\alpha\beta}\,\theta_{\mu\nu}
+\left({\not\!  \partial}\right)_{\alpha\beta}\, V_\mu
\nonumber\\[3mm]
&+&\varepsilon_{\mu\nu}\,\partial^\nu
\left[ A\, \delta_{\alpha\beta}\,  R_{i\bar j}\bar\Psi^{\bar j}\gamma^5 \Psi^{i}
+ B \left(\gamma^5\right)_{\alpha\beta}\, R_{i\bar j}\bar\Psi^{\bar j} \Psi^{i}
\right],
\label{genfac}
\eeqn
where $V_\mu$ is a conserved vector current,
$\mu$ and $\nu$ are vectorial indices while $\alpha$ and $\beta$
are spinor indices, $A$ and $B$ are dimensioneless numbers.
Equation (\ref{genfac}) also uses dimensional arguments,
the target space covariance and the fact that
the central charges (the second line in Eq. (\ref{genfac}))
do not appear at the classical level. This leaves us with two dimensionelss constants,
$A$ and $B$, to be determined. Now, we convolute both sides with
$ \gamma^\mu $ to get
\beqn
&&\left\{\bar Q _\beta, \,\, \frac{R_{i\bar j}}{2\pi}\,
\left(\partial_\lambda \bar\phi^{\, \bar j} \right) \left(\gamma^\lambda \Psi^i 
\right)_\alpha \right\} =
\delta_{\alpha\beta}\,\theta_{\mu}^{\mu}
+\left(\gamma^5\right)_{\alpha\beta}\, \partial_\mu A^\mu
\nonumber\\[3mm]
&&+\varepsilon_{\mu\nu}\,\partial^\nu
\left[ A\, \left(\gamma^\mu\right)_{\alpha\beta}\,  R_{i\bar j}\bar\Psi^{\bar j}\gamma^5 \Psi^{i}
+ B \left(\gamma^\mu\gamma^5\right)_{\alpha\beta}\, R_{i\bar j}\bar\Psi^{\bar j} \Psi^{i}
\right]\, ,
\label{convo}
\eeqn
where $ A^\mu$ is an axial current.
We perform the supertransformation on the left-hand side.
Then, to  untangle the $A$  
structure we multiply by $\left(\gamma^\rho\right)_{\beta\alpha}$ 
and to  untangle the $B$  
structure by $\left(\gamma^\rho\gamma^5\right)_{\beta\alpha}$. 
After simple algebra 
we get
\beqn
\{\bar  Q_R \,,\,  Q_L \} &=&
\frac{1}{2 \pi}  \int dx\, \partial_x \left( R_{i\bar j}\,
\bar\Psi_R^{\bar j} \Psi_L^i\right) \,,
\nonumber\\[3mm]
\{\bar  Q_L \,,\,  Q_R \}
&=& \frac{1}{2\pi}  \int dx\, \partial_x \left( R_{i\bar j}\,
\bar\Psi_L^{\bar j} \Psi_R^i\right) \,.
\label{twentythree}
\eeqn
Although these expressions are obtained
in the round metric, we conjecture that in this form they
are valid for other K\"ahler sigma models too.

 As well-known,
the models at hand possess a discrete set of supersymmetric vacua
labeled by the order parameters
$ \langle R_{i\bar j}\,
\bar\Psi_R^{\bar j} \Psi_L^i\rangle $ and $ \langle  R_{i\bar j}\,
\bar\Psi_L^{\bar j} \Psi_R^i \rangle $. Thus,
the   central charges (\ref{twentythree})
determine BPS soliton masses. We plan to discuss
the issue in more detail elsewhere.

\section{Conclusions}
\label{conclu}
\setcounter{equation}{0}

We have considered a new class of two-dimensional ${\cal N}=2$
sigma models, with twisted mass {\em and } a
superpotential. The requirement of ${\cal N}=2$ supersymmetry uniquely
fixes the form of the superpotential.
The emerging theory is rather peculiar --- its superalgebra
contains
central  charges in the anticommutators
$\{ Q_L, Q_L\}$ and $\{ Q_R, Q_R\}$
which are proportional to the spatial size of the system
and 
 have no parallels in the previous investigations
of supersymmetric field theories.
Due to the occurence of these central charges
the vacuum energy density does not vanish,
and yet 1/2 of supersymmetry is realized linearly.
The vacua can be viewed as 1/2 BPS saturated states.
This result defies a well-known theorem
prohibiting spontaneous breaking of a part of extended supersymmetry.

In the model we deal with the standard fermion number is not defined,
so that   the Witten index as well as the
Cecotti-Fendley-Intriligator-Vafa index  
are not applicable. 
We suggest an alternative definition of an index  (in internal algebraic terms)
that counts short multiplets. Our definition
works 
 in spite of the absence of the standard fermion number.

In the last section, we outlined a derivation of   quantum anomaly in
the anticommutator $\{ \bar Q_L, Q_R\}$ (in the undeformed theory,
with no superpotential and twisted mass).
An anomalous central charge emerging in this
anticommutator is proportional to
$\Delta \langle   R_{i\bar j}\,
\bar\Psi_L^{\bar j} \Psi_R^i \rangle $.
The expression in $\langle ... \rangle$ is the order parameter
known to take different (known) values
in distinct vacua. Therefore, the critical soliton masses
are related to this order parameter.
On the other hand, these masses can be calculated
by using other methods. It seems very interesting to compare
these dual calculations. This will be a subject of a separate
investigation.

\section*{Acknowledgments}

We are grateful to E.~A.~Ivanov and S.~J.~Gates  for communications.
Special thanks go to A. Vainshtein for insightful discussions.

The work of A.L. is supported in part by Theoretical 
Physics Institute at the University of Minnesota,
by the RFFI grant 01-01-00548,
by the INTAS grant 99-590,
and by the grant for Support of the Scientific School of M.A. Olshanetsky.
The work of M.S. is supported in part by the DOE grant
DE-FG02-94ER408.

\end{document}